\documentclass[
    ,final            
]
  {aipproc}

\layoutstyle{6x9}

\usepackage{graphics,epsfig}
\usepackage{amsmath}
\usepackage{graphicx}
\usepackage{amssymb}
\usepackage{amsmath}
\usepackage{latexsym}

\newcommand{\be}{\begin{equation}}
\newcommand{\ee}{\end{equation}}
\newcommand{\bq}{\begin{eqnarray}}
\newcommand{\eq}{\end{eqnarray}}

\newcommand{\D}{\mathrm{d}}

\newcommand{\nslash}{\kern 0.2 em n\kern -0.50em /}
\newcommand{\kslash}{\kern 0.2 em k\kern -0.45em /}
\newcommand{\pslash}{\kern 0.2 em p\kern -0.50em /}
\newcommand{\Sslash}{\kern 0.2 em S\kern -0.50em /}
\newcommand{\Pslash}{\kern 0.2 em P\kern -0.50em /}
\newcommand{\Dslash}{\kern 0.2 em D\kern -0.65em /\kern 0.15em}

\def\lsim{\mathrel{\rlap{\lower4pt\hbox{\hskip1pt$\sim$}}\raise1pt\hbox{$<$}}}
\def\gsim{\mathrel{\rlap{\lower4pt\hbox{\hskip1pt$\sim$}}\raise1pt\hbox{$>$}}}
\def\Vec#1{\mathpalette{\VVec}{#1}}                  
\def\VVec#1#2{\mbox{\boldmath$#1#2$\unboldmath}}
\def\nostrocostruttino#1\over#2{\mathrel{\mathop{\kern 0pt \rlap
{\hbox{$#1$}}} \hbox{\kern-.135em $#2$}}}

\def\anti#1{\mathpalette{\@anti}{#1}#1}
\def\@anti#1#2{\sbox0{$#1#2$}
\makebox[0pt][l]{$#1\kern.30\ht0\overline{\kern-.35\ht0\phantom{#2}}$}}

\begin{document}

\title{SIDIS in target fragmentation region }


\author{\underline {A.~Kotzinian}}{
  address={\it Dipartimento di Fisica Teorica, Universit{\`a}
di Torino; \\
INFN, Sezione di Torino, 10125 Torino, Italy}
,altaddress={\it Yerevan Physics Institute, 375036 Yerevan, Armenia} 
}

\author{M.~Anselmino}{
  address={\it Dipartimento di Fisica Teorica, Universit{\`a}
di Torino; \\
INFN, Sezione di Torino, 10125 Torino, Italy}
}

\author{V.~Barone}{
  address={\it Di.S.T.A., Universit{\`a} del Piemonte
Orientale ``A. Avogadro''; \\
INFN, Gruppo Collegato di Alessandria,  15121 Alessandria, Italy}}

\begin{abstract}

We shortly describe the leading twist formalism of spin and transverse-momentum dependent fracture functions recently developed and present results for the production of spinless hadrons in the target fragmentation region (TFR) of
SIDIS~\cite{Anselmino:2011ss}. In this case not all fracture functions can be accessed and only a Sivers-like single spin azimuthal asymmetry shows up at LO cross-section.
Then, we demonstrate that the measurement of spin dependent azimuthal asymmetries in double hadron production in polarized SIDIS -- with one spinless hadron produced in the current fragmentation region (CFR) and another in the TFR -- would  provide access to all 16 leading twist fracture functions.

\end{abstract}

\maketitle

\section{Introduction}

As it is becoming increasingly clear in the last decades, the study of the
three-dimensional spin-dependent partonic structure of the nucleon in SIDIS processes requires a full understanding of the hadronization process after the hard lepton-quark scattering. So far most SIDIS experiments were studied in the CFR, where an adequate theoretical formalism based on distribution and fragmentation functions has been established, see for example~\cite{Bacchetta:2006tn}. However, to avoid
misinterpretations, also the factorized approach to SIDIS description in the TFR
has to be explored. The corresponding theoretical basis -- the fracture functions formalism -- was established in Ref.~\cite{Trentadue:1993ka}
for hadron transverse momentum integrated unpolarized cross-section. Recently this approach was generalized~\cite{Anselmino:2011ss} to the spin and transverse momentum dependent case (STMD).

We consider the process
\begin{equation}\label{sidis-tfr}
\ell(l,\lambda)+N(P_N,S) \to \ell(l')+h(P)+X
\end{equation}
with the hadron $h$ produced in the TFR. We use the standard DIS notations and in the $\gamma^*-N$ c.m. frame we define the $z$-axis along the direction of
${\bf q}$ (the virtual photon momentum) and the $x$-axis along ${\bf l}_T$, the  lepton transverse momentum. The kinematics of the produced hadron is defined by
its transverse momentum $P_T$ and azimuthal angle $\phi$. Assuming TMD factorization the cross-section of the process (\ref{sidis-tfr}) can be written as
\begin{equation}\label{sidis-tfr-cs}
\frac{d\sigma^{\ell(l,\lambda)+N(P_N,S) \to \ell(l')+h(P)+X}}
{dx \, dQ^2\, d\phi_S \, d \zeta \, d^2P_{T}} =
{\cal M}_{q,s/N,S}^{h} \otimes \frac{d\sigma^{\ell(l,\lambda)+q(k,s) \to \ell(l')+q(k',s')}}{dQ^2}\,,
\end{equation}
where the STMD fracture functions ${\cal M}_{q,s/N,S}^h$ has a clear probabilistic meaning: it is the conditional probability to produce a hadron $h$ in the TFR when the hard scattering occurs on a quark $q$ from the target nucleon $N$. The expression of the non-coplanar polarized lepton-quark hard scattering cross-section can be found in Ref.~\cite{Kotzinian:1994dv}.

The most general expressions of the LO STMD fracture functions for unpolarized, longitudinally polarized ($s_L$) and transversely polarized (${\bf s}_T$) quarks can be written as~\cite{Anselmino:2011ss}:
\begin{eqnarray}
M_{q/N,S}^{h}
&=&
\hat{ M}
+ \frac{ {\bf P}_T \times  {\bf S}_T}{m_h} \,
\hat{M}_T^h + \frac{ {\bf k} \times
 {\bf S}_T}{m_N} \, \hat{M}_T^{\perp}
 +
\frac{S_L \, ( {\bf k}
\times  {\bf P}_T)}{m_N \, m_h} \,
\hat{M}_L^{\perp h}
\label{up-frf} \\
\Delta M_{q,s_L/N,S}^{h} & = & \hat{ M} s_L=
S_L \, \Delta \hat{M}_L
+ \frac{  {\bf P}_T \cdot  {\bf S}_T}{m_h} \,
\Delta \hat{M}_T^h
+ \frac{ {\bf k} \cdot  {\bf S}_T}{m_N}
\, \Delta \hat{M}_T^{\perp}
 +  \frac{ {\bf k} \times
 {\bf P}_T}{m_N \, m_h} \,
\Delta \hat{M}^{\perp h}
\label{lp-frf} \\
\Delta_T M_{q,{\bf s}_T/N,S}^{h_2} & = & \hat{ M} s_T^i
= S_T^i \, \Delta_T \hat{M}_T
+ \frac{S_L \, P_T^i}{m_h} \, \Delta_T \hat{M}_L^h
+ \frac{S_L \, k^i}{m_N} \,
\Delta_T \hat{M}_L^{\perp}
\nonumber \\
& & + \, \frac{( {\bf P}_T \cdot  {\bf S}_T)
\, P_T^i}{m_h^2} \, \Delta_T \hat{M}_T^{hh}
+ \frac{( {\bf k} \cdot  {\bf S}_T)
\, k^i}{m_N^2} \, \Delta_T \hat{M}_T^{\perp \perp}
\nonumber \\
& & + \frac{( {\bf k} \cdot  {\bf S}_T)
\, P_T^i - ( {\bf P}_T \cdot  {\bf S}_T)
\, k^i }{m_N m_h} \, \Delta_T \hat{M}_T^{\perp h}
\nonumber \\
& & + \, \frac{\epsilon_{\perp}^{ij} P_{T j}}{m_h}
\, \Delta_T \hat{M}^h
+ \frac{\epsilon_{\perp}^{ij} k_j}{m_N}
\, \Delta_T \hat{M}^{\perp}\,,
\label{tp-frf}
\end{eqnarray}
where ${\bf k}$ is the quark transverse momentum and by the vector product of the two-dimensional vectors  $ {\bf a}$ and ${\bf b}$ we mean the
pseudo-scalar quantity $ {\bf a} \times  {\bf b} =
\epsilon^{i j} \, a_i b_j =  a   b  \, \sin (\phi_b - \phi_a)$.
All fracture functions depend on the scalar variables
$x,  k^2, \zeta$ (defined in the sequel), $P_T^2$ and ${\bf k} \cdot  {\bf P}_T$. For the production of a spinless hadron in the TFR one has~\cite{Anselmino:2011ss}:
\bq
& & \frac{d\sigma^{\ell(l,\lambda)+N(P_N,S) \to \ell(l')+h(P)+X}}
{dx \, dQ^2\, d\phi_S \, d \zeta \, d^2P_{T}}
 =
\frac{\alpha_{\rm em}^2}{Q^2 y} \,
\nonumber \\
&&
\hspace{-0.5cm}
\times \Bigg \{
\left (1 +(1-y)^2 \right )
\, \sum_a e_a^2 \,
 \left  [   M(x, \zeta, P_T^2)
-  \Vec S_T \, \frac{P_T}{m_h}
\, M_{T}^h (x, \zeta, P_T^2) \, \sin (\phi_h - \phi_S)
\right ]
\nonumber \\
& &
\hspace{-0.5cm} + \,
\lambda \, y \, (2 - y )
\sum_a e_a^2 \,
 \left [
S_L \, \Delta M_{L} (x, \zeta, P_T^2)
+ \, S_T\, \frac{P_T}{m_h}
\, \Delta M_{T}^h (x, \zeta, P_T^2) \, \cos (\phi_h - \phi_S)
\right ] \Bigg \}\,,
\label{cross1}
\eq
where the ${\bf k}$-integrated fracture functions are given as
\be
M(x, \zeta, P_T^2)
= \int \D^2 k\,
\hat{M} \, ,
\quad
M_T^h(x, \zeta, P_T^2)
= \int \D^2 k \,
\left \{ \hat{M}_T^h
+ \frac{m_h}{m_N}
\frac{{\bf k} \cdot {\bf P}_T}{P_T^2}
\, \hat{M}_T^{\perp} \right \}\, ,
\label{intm2}
\ee
\be
\Delta M_L(x, \zeta, P_T^2)
= \int \D^2 k \,
\Delta \hat{M}_L \,,
\quad
\Delta M_T^h (x, \zeta, P_T^2)
= \int \D^2 k
\, \left \{ \Delta \hat{M}_T^h +
\frac{m_h}{m_N}
\frac{{\bf k} \cdot {\bf P}_T}{P_T^2}
\, \Delta \hat{M}_T^{\perp} \right \}\,.
\label{intdeltam2}
\ee

We see that a single hadron production in the TFR of SIDIS does not provide access to all fracture functions. At LO the cross-sections contains only the Sivers-like single spin azimuthal asymmetry.

\section{DSIDIS}
In order to have access to all fracture functions one has to "measure" the scattered quark transverse polarization, for example exploiting he Collins effect~\cite{Collins:1992kk}, which gives the probability to find an unpolarized hadron in the jet originated from a transversely polarized quark:
\be
D_{q,s'}^{h}(z,{\bf p})=D_1(z,p^2)+\frac{ {\bf p} \times  {\bf s}'_T}{m_h}H_1^\perp(z,p^2)\, ,
\ee
where ${\bf p}$ is the transverse momentum of the hadron with respect to the fragmenting quark momentum and ${\bf s}'_T$ is the transverse quark polarization.

\begin{figure}[h!]
\label{fig:sidis-assoc}
  \includegraphics[height=.19\textheight]{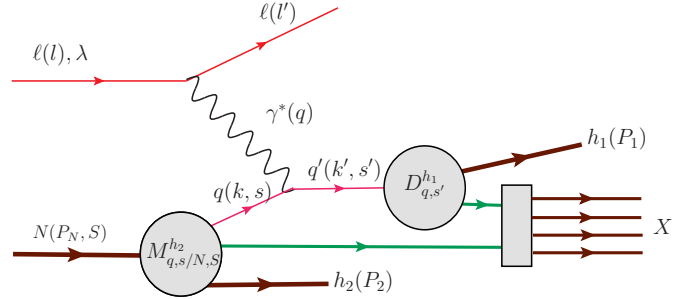}
  \caption{DSIDIS description in factorized approach at LO.}
\end{figure}

Let us consider a double hadron production process (DSIDIS)
\begin{equation}\label{dsidis}
\ell(l)+N(P_N) \to \ell(l')+h_1(P_1)+h_2(P_2)+X
\end{equation}
with (unpolarized) hadron 1 produced in the CFR ($x_{F1}>0$) and hadron 2 in the TFR ($x_{F2}<0)$), see Fig. \ref{fig:sidis-assoc}.  For hadron $h_1$ we will use the ordinary scaled variable $z = P_1^+/k'^+ \simeq P_N{\cdot}P_1/P_N{\cdot}q$, its transverse momentum $P_{T1}$ and azimuthal angle $\phi_1$ and for hadron $h_2$
the variables $\zeta = P_2^-/P_N^- \simeq E_1/E_N$, $P_{T2}$ and $\phi_2$.

In this case the LO expression for the DSIDIS cross-section includes all fracture function:
\begin{eqnarray}\label{cs-2h}
\frac{d\sigma^{\ell(l,\lambda)+N(P_N,S) \to \ell(l')+h_1(P_1)+h_2(P_2)+X}}
{dx \, dy\, d\phi_S \,dz\, d^2P_{T1}\,d \zeta\, d^2P_{T2}}
& = &
\nonumber \\
&&
\hspace{-8cm}
\frac{\alpha^2\,x}{Q^4 y}\left(1+(1-y)^2\right)
\bigg(M_{q/N,S}^{h_2} \otimes D_{1q}^{h_1}+
\lambda D_{ll}(y)\Delta M_{q,s_L/N,S}^{h_2} \otimes D_{q}^{h_1}
+
D_{nn}(y)\Delta_T M_{q,{\bf s}_T/N,S}^{h_2} \otimes \frac{{\bf p} \times  {\bf s}'_T}{m_{h_1}}H_{1q}^{\perp h_1}
\bigg) =
\nonumber \\
&&
\hspace{-7cm}
\frac{\alpha^2\,x}{Q^4 y}\left(1+(1-y)^2\right)
\left(\sigma_{UU}+S_L\,\sigma_{UL}+S_T\,\sigma_{UT}+
\lambda \,D_{ll} \sigma_{LU}+\lambda \,S_L D_{ll}\,\sigma_{LL}+\lambda \,S_T D_{ll}\,\sigma_{LT} \right)\, ,
\end{eqnarray}
where
\begin{equation}\label{D_ll}
D_{ll}(y) = \frac{y(2-y)}{1+(1-y)^2}\,,\qquad D_{nn}(y) = \frac{2(1-y)}{1+(1-y)^2}\, .
\end{equation}

We show here explicit expressions only for the case of unpolarized target\footnote{Expressions for other terms are available in~\cite{Kotzinian:DIS2011}.}
\begin{eqnarray}\label{s_uu}
\sigma_{UU} & = & F_0^{{\hat M} \cdot D_1}-D_{{nn}} \Bigg[\frac{P_{{T1}}^2 }{m_1 m_N}\, F_{{kp1}}^{\Delta_T {\hat M}^\perp \cdot H_1^\perp}\,{\cos}(2 \phi _1)
+ \frac{P_{{T1}} P_{{T2}} }{m_1 m_2}\, F_{{p1}}^{\Delta_T {\hat M}^h \cdot H_1^\perp}\, {\cos}(\phi _1+\phi _2)
\nonumber \\
& + & \left(\frac{P_{{T2}}^2 }{m_1 m_N}\, F_{{kp2}}^{\Delta_T {\hat M}^\perp \cdot H_1^\perp} + \frac{P_{{T2}}^2 }{m_1 m_2}\, F_{{p2}}^{\Delta_T {\hat M^h} \cdot H_1^\perp}\right)\, {\cos}(2 \phi _2)\Bigg].
\nonumber \\
\sigma_{LU} & = &-\frac{ P_{{T1}} P_{{T2}}}{m_2 m_N}  F_{{k1}}^{\Delta{\hat M}^{\perp h}\cdot D_1} \, \sin(\phi _1-\phi _2)
\, ,
\end{eqnarray}
where the structure functions $F_{...}^{...}$ are specific convolutions~\cite{Kotzinian:DIS2011} of fracture and fragmentation functions depending on $x, \zeta, P_{T1}^2,  P_{T2}^2, {\bf P}_{T1} \cdot  {\bf P}_{T2}$.

We notice the presence of terms similar to the Boer-Mulders term appearing in the usual CFR of SIDIS. What is new in DSIDIS is the LO beam spin SSA, absent in the CFR of SIDIS.
We further notice that the DSIDIS structure functions may depend in principle on the relative azimuthal angle of the two hadrons, due to presence of the last term among their arguments:
${\bf P}_{T1} \cdot {\bf P}_{T2} = P_{T1} P_{T2}\cos(\phi_1-\phi_2)$. This term arise from ${\bf k} \cdot  {\bf P}_T$ correlation in STMD fracture functions and can generate a long range correlations between hadrons produced in CFR and TFR.
In practice these correlations will complicate the analysis of the azimuthal dependence of cross-sections. For example, expanding $F_{{k1}}^{\Delta{\hat M}^{\perp h}\cdot D_1}$ in series of ${\bf P}_{T1} \cdot {\bf P}_{T2}$ one obtains $\sigma_{LU} \propto a_1\sin(\Delta \phi)+a_2\sin(2\Delta \phi)+\cdots$, with $\Delta \phi=\phi_1-\phi_2$ and amplitudes $a_1,a_2,\dots$ independent of azimuthal angles.

We stress that the ideal occasions to test the predictions of the present approach to DSIDIS, would be the future JLab 12 upgrade, in progress, and the EIC facilities, in the planning phase.

\bibliographystyle{aipproc}

\end{document}